\documentclass[12pt,prc]{article}
\usepackage{hyperref}
\usepackage[utf8]{inputenc}
\RequirePackage{lineno}
\usepackage[english]{babel}
\usepackage{xcolor}
\usepackage{multirow}
\usepackage{graphicx}
\usepackage{dcolumn}
\usepackage{amsmath,mathtools}
\usepackage{supertabular}
\usepackage{multirow}
\RequirePackage{lineno}
\usepackage{color}
\usepackage{cancel}
\usepackage{ulem}
\usepackage{array}
\usepackage{tabularx}
\usepackage[T1]{fontenc}
\usepackage{amsmath}
\usepackage{authblk}

\newcommand{\nuStatTwo}{$\nu_{\pm,\rm stat}^{\rm 2nd}$}
\newcommand{\nuStatThree}{$\nu_{\pm,\rm stat}^{\rm 3rd}$}
\newcommand{\nuStatFour}{$\nu_{\pm, \rm stat}^{\rm 4th}$}
\newcommand{\nudynTwo}{$\nu_{\pm,dyn}^{\rm 2nd}$}
\newcommand{\nudynThree}{$\nu_{\pm,dyn}^{\rm 3rd}$}
\newcommand{\nudynFour}{$\nu_{\pm,dyn}^{\rm 4th}$}
\newcommand{\nudynM}{$\nu_{\pm,dyn}^{m}$}
\newcommand{\sNN}{$\sqrt{s_{NN}}$}
\newcommand{\pT}{$p_{\rm T}$}

\title{\textbf{Higher order dynamical charge fluctuations in heavy-ion collisions}}
\author{Bhanu Sharma and P. K. Sahu }
\affil{Institute of Physics, HBNI, Bhubaneswar 751005, India}
\date{\today}
\begin{document}

\maketitle
\begin{abstract}
The event-by-event charge fluctuation measurements are proposed to provide the signature of quark-gluon plasma (QGP) in heavy-ion collisions. Measure of dynamical charge fluctuations is expected to carry information of initial fractional charge of the QGP phase at the final state. We propose the higher order charge fluctuation measurement to study the QGP signal in heavy-ion collisions. This higher order charge fluctuation observable can amplify the signature of QGP. Also, the SMASH model is used to study the behavior of these observable in heavy-ion collisions at center of mass energies accessible in the STAR beam energy scan program. 
\end{abstract}
\maketitle

\section{Introduction}
\indent Shortly after the Big Bang, the deconfined state of Quantum Chromodynamics (QCD) matter--known as the Quark-Gluon Plasma (QGP)--filled a few microsecond old universe. The dedicated facilities at Large Hadron Collider (LHC) and the Relativistic Heavy Ion Collider (RHIC) have been trying to recreate the conditions very similar to the early universe by making head-on collisions between heavy-ions ~\cite{STAR_2005,PHENIX:2004vcz,BRAHMS:2004adc,ALICE:2008ngc,CMS:2008xjf}. 
 
 In heavy-ion collisions, the transition from the QGP to normal hadron gas occurs at chemical freeze out in a short time scale. However, the QGP phase carries fractional charge due to the presence of quarks and the hadron resonance gas (HRG) phase contains the integral charge of hadrons. Hence, it is expected that the charge fluctuations measurement may carry the information of initial fractional charge to the final state and can be measured at the detectors.

 The main idea is that the charge fluctuations are directly proportional to the square of charges present in the system. Due to the  difference between the charges carried by quarks in QGP as compared to hadrons in hadronic phase, a dramatic reduction of event-by-event fluctuations of the charge has been predicted by various theoretical calculations ~\cite{Jeon:2000wg,Heiselberg:2000ti,Asakawa:2000wh,Shuryak:2000pd,Jeon:1999gr,Bleicher:2000ek}.

 As of now, the charge fluctuations have been analysed by various experiments. The first measurements were done by PHENIX ~\cite{PHENIX:2002eqg} and STAR ~\cite{STAR:2003oku} for Au+Au collisions at $\sqrt{s_{NN}}$ = 130 GeV. The charge fluctuations by PHENIX experiment were measured (observable in ~\cite{PHENIX:2002eqg} ) and the results are in qualitative agreement with HRG calculation and far away from the plasma. The measurements reported by STAR experiment were in terms of a robust variable, known as dynamical charge fluctuations. The behaviour of dynamical charge fluctuations was studied ~\cite{STAR:2008szd} with varying beam energy and different collision system size like Au+Au and Cu+Cu collisions. However, the measurements were in agreement with the previous measurements. The charge ﬂuctuations measured by ALICE experiment ~\cite{ALICE:2012xnj} for Pb+Pb collisions at $\sqrt{s_{NN}}$ = 2.76 TeV for $\Delta \eta$ = 1.0 are closer to the theoretical predictions for the QGP formation. Recently, a similar measurement by the CMS experiment has been reported for charge fluctuations in Pb+Pb collisions at $\sqrt{s_{NN}}$ = 5.02 TeV for a wide pseudorapidity range $\Delta \eta$ = 4.8 ~\cite{CMS:2023drv}. The measured value reaches the QGP predictions even without considering the global charge conservation in account. This indicates that to measure a possible signal for QGP formation, we need to measure at very high energies and large acceptance range. 
 
 In this calculation, we propose the higher orders of dynamical charge fluctuations measure in heavy-ion collisions that can amplify the signature of QGP within limited kinematic phase space.
 The paper is organized into various sections. In Section~\ref{Sect:MotivAndObser}, the derivation of the observable and methodology are introduced. In Section~\ref{Sect:SMASHmodel}, the event-generation model description is mentioned that is used to study these observables. In Section~\ref{Sect:SimAna}, simulation inputs and analysis details are discussed. Section~\ref{Sect:resultsAndDis} presents the results with discussion. Finally, Section~\ref{Sect:SummaryOutlook} summarizes and outlooks this study.

\section{Motivation and obvervables}
\label{Sect:MotivAndObser}

 The observable of dynamical charge fluctuations, $\nu_{\pm,dyn}$, has already been introduced and measured in heavy-ion experiments~\cite{STAR:2003oku,STAR:2008szd,ALICE:2012xnj,CMS:2023drv}. A detailed discussion on this observable is mentioned in Sec.~\ref{Sect:SecondOrder}. However, one can readily extend this observable to their $m^{\rm th}$ order as,
\begin{equation}
\label{Eq:mthOrder}
    \nu_{\pm}^{\mathrm{m}} =\Bigg \langle \Bigg(\frac{N_{+}}{\langle N_{+}\rangle} - \frac{N_{-}}{\langle N_{-}\rangle}\Bigg)^{m}\Bigg \rangle.
\end{equation}
 Here, $m$ is an positive integer. $N_{\pm}$ are the event-by-event number of positive or negative charged particles. The $<\dots>$ represents ensemble average. By definition, the  1st order of this observable vanishes, whereas higher order observables contain higher order correlations terms between positive and negative charged particles per event.  The $2^{\rm nd}$, $3^{\rm rd}$, and $4^{\rm th}$ order of these observables are discussed in the following subsections.   

\subsection{\textbf{Second order dynamical charge fluctuations}}
\label{Sect:SecondOrder}
The $2^{\rm nd}$ order of Eq.~\ref{Eq:mthOrder} can be written as,
	\begin{equation}
 \begin{split}  
 \label{Eq:Eqnu2nd}
	\nu_{\pm}^{\mathrm{2nd}} &=\Bigg \langle \Bigg(\frac{N_{+}}{\langle N_{+}\rangle} - \frac{N_{-}}{\langle N_{-}\rangle}\Bigg)^{2}\Bigg \rangle \\
	&=\frac{\langle N_{+}^2\rangle}{\langle N_{+}\rangle^2}+\frac{\langle N_{-}^2\rangle}{\langle N_{-}\rangle^2}-2\frac{\langle N_{+}N_{-}\rangle}{\langle N_{+}\rangle\langle N_{-}\rangle}.
 \end{split}
	\end{equation}
 The above equation contains cross-correlation between $N_{+}$ and  $N_{-}$. 
 
 For independent particle production, the cross-correlation becomes uncorrelated and the statistical contribution becomes :
 	\begin{equation}
	 \label{Eq:Eqnustat2}
\nu_{\pm,\mathrm{stat}}^{\mathrm{2nd}}=\frac{1}{\langle N_{+}\rangle}+\frac{1}{\langle N_{-}\rangle}.
	\end{equation}
 
 Hence, the dynamical charge fluctuations of $2^{\rm nd}$ order between $N_{+}$ and $N_{-}$ can be expressed using Eq.~\ref{Eq:Eqnu2nd} and ~\ref{Eq:Eqnustat2} as follows,
 \begin{equation}
 \label{Eq:Eqnudyn2}
  \begin{split} 
	\nu_{\pm,dyn}^{\mathrm{2nd}} &=\nu_{\pm}^{\mathrm{2nd}}-\nu_{\pm,\mathrm{stat}}^{\mathrm{2nd}} \\
	&= \frac{\langle N_{+}(N_{+}-1)\rangle}{\langle N_{+}\rangle^2}+\frac{\langle N_{-}(N_{-}-1)\rangle}{\langle N_{-}\rangle^2}-2\frac{\langle N_{+}N_{-}\rangle}{\langle N_{+}\rangle \langle N_{-}\rangle}.
\end{split}
 \end{equation}

 One can express the above equation in terms of factorial moments; in factorial moments the detection efficiencies of $N_{\pm}$ are factorized with an ansatz of the Binomial detector response. The $\nu_{\pm,dyn}^{\mathrm{2nd}}$ is a robust variable as it has no detector efficiency effect~\cite{Pruneau:2002yf}.

\subsection{\textbf{Third order dynamical charge fluctuations}}
\label{Sect:3rdOrdernu}
The $3^{\rm rd}$ order of Eq.~\ref{Eq:mthOrder} can be written as,
 	\begin{equation}
    \label{Eq:3rdnu}
    \begin{split}    
 \nu_{\pm}^{\mathrm{3rd}} &=\Bigg \langle \Bigg(\frac{N_{+}}{\langle N_{+}\rangle} - \frac{N_{-}}{\langle N_{-}\rangle}\Bigg)^{3}\Bigg \rangle \\
	&=\frac{\langle N_{+}^3\rangle}{\langle N_{+}\rangle^3}-\frac{\langle N_{-}^3\rangle}{\langle N_{-}\rangle^3}-3\frac{\langle N_{+}^2 N_{-}\rangle}{\langle N_{+}\rangle^2 \langle N_{-}\rangle}+3\frac{\langle N_{+} N_{-}^2\rangle}{\langle N_{+}\rangle \langle N_{-}\rangle^2}.
  \end{split}
	\end{equation}
For independent particle production and considering both $N_{\pm}$ multiplicity distributions are the Poisson distributions, the statistical contribution of $3^{\rm rd}$ order becomes :
\begin{equation}
\nu_{\pm,\mathrm{stat}}^{\mathrm{3rd}}	=\frac{1}{\langle N_{+}\rangle^2}-\frac{1}{\langle N_{-}\rangle^2}.
	\end{equation}
 Using above equations, the dynamical charge fluctuation of $3^{\rm rd}$ order can be expressed as : 
 \begin{equation}
 \label{Eq:Eq3rdnudyn}
 \begin{split}    
	\nu_{\pm,dyn}^{\mathrm{3rd}} &=\nu_{\pm}^{\mathrm{3rd}}-\nu_{\pm,\mathrm{stat}}^{\mathrm{3rd}}\\
	& = \frac{\langle N_{+}^3\rangle}{\langle N_{+}\rangle^3}-\frac{\langle N_{-}^3\rangle}{\langle N_{-}\rangle^3}-3\frac{\langle N_{+}^2 N_{-}\rangle}{\langle N_{+}\rangle^2 \langle N_{-}\rangle}+3\frac{\langle N_{+} N_{-}^2\rangle}{\langle N_{+}\rangle \langle N_{-}\rangle^2}-\frac{1}{\langle N_{+}\rangle^2}+\frac{1}{\langle N_{-}\rangle^2}.
  \end{split}
	\end{equation}
 Here it is worth to note here that $\nu_{\pm,dyn}^{\mathrm{3rd}}$ contains higher order cross-correlations between $N_{+}$ and $N_{-}$ and also the last two terms just remains unfactorized in terms of factorial moments. Hence, $\nu_{\pm,dyn}^{\mathrm{3rd}}$ is not a robust observable for detector efficiency effect in the experiment.

	\subsection{\textbf{Fourth order dynamical charge fluctuations}}
 \label{Sect:Sec4nudyna}
 
 The $4^{\rm th}$ order of Eq.~\ref{Eq:mthOrder} can be written as,
 
 \begin{equation}
 \label{Eq:Eq4thnu}
 \begin{split}    
	\nu_{\pm}^{\mathrm{4th}} &=\Bigg \langle \Bigg(\frac{N_{+}}{\langle N_{+}\rangle} - \frac{N_{-}}{\langle N_{-}\rangle}\Bigg)^{4}\Bigg \rangle \\
	&=\frac{\langle N_{+}^4\rangle}{\langle N_{+} \rangle^4}+\frac{\langle N_{-}^4\rangle}{\langle N_{-} \rangle^4}-4\frac{\langle N_{+}^3 N_{-}\rangle}{\langle N_{+}\rangle^3 \langle N_{-}\rangle}-4\frac{\langle N_{+}N_{-}^3\rangle}{\langle N_{+}\rangle \langle N_{-}\rangle^3}+6\frac{\langle N_{+}^2 N_{-}^2\rangle}{\langle N_{+}\rangle^2 \langle N_{-}\rangle^2}.
 \end{split}    
	\end{equation}
 For independent particle production and considering both $N_{\pm}$ multiplicity distributions are the Poisson distributions, the statistical contribution of $4^{\rm th}$ order becomes :

  \begin{equation}
  \label{Eq:Eq4nustat}
\nu_{\pm,\mathrm{stat}}^{\mathrm{4th}}	=\frac{1}{\langle N_{+}\rangle^3}+\frac{1}{\langle N_{-}\rangle^3}+\frac{3}{\langle N_{+}\rangle^2}+\frac{3}{\langle N_{-}\rangle^2}+\frac{6}{\langle N_{+}\rangle \langle N_{-}\rangle}.
\end{equation}
Using above equations, the dynamical charge fluctuation of $4^{\rm th}$ order can be expressed as : 
\begin{equation}
\label{Eq:Eq4thdyn}
\begin{split}
  \nu_{\pm,dyn}^{4th} &= \nu_{\pm}^{\mathrm{4th}}-\nu_{\pm,\mathrm{stat}}^{\mathrm{4th}}\\
    &= \!\begin{multlined}[t]
     \frac{\langle N_{+}^4\rangle}{\langle N_{+} \rangle^4}+\frac{\langle N_{-}^4\rangle}{\langle N_{-} \rangle^4}- 4\frac{\langle N_{+}^3 N_{-}\rangle}{\langle N_{+}\rangle^3 \langle N_{-}\rangle}-4\frac{\langle N_{+}N_{-}^3\rangle}{\langle N_{+}\rangle \langle N_{-}\rangle^3}+
 6\frac{\langle N_{+}^2 N_{-}^2\rangle}{\langle N_{+}\rangle^2 \langle N_{-}\rangle^2}\\ -\frac{1}{\langle N_{+}\rangle^3}-
 \frac{1}{\langle N_{-}\rangle^3}-\frac{3}{\langle N_{+}\rangle^2}-\frac{3}{\langle N_{-}\rangle^2}-\frac{6}{\langle N_{+}\rangle \langle N_{-}\rangle}.
     \end{multlined}
     \end{split}
\end{equation}
 Here the $\nu_{\pm,dyn}^{\mathrm{4th}}$ contains higher order cross-correlations between $N_{+}$ and $N_{-}$ and also several terms just remains unfactorized in terms of factorial moments. However, unlike $\nu_{\pm,dyn}^{\mathrm{2nd}}$, the $\nu_{\pm,dyn}^{\mathrm{3rd}}$ and $\nu_{\pm,dyn}^{\mathrm{4th}}$ are not  robust observables for detector efficiency effect in the experiment and hence the $\nu_{\pm,dyn}^{\mathrm{3rd}}$ and $\nu_{\pm,dyn}^{\mathrm{4th}}$ need additional detector effect correction in experiment. 
 
 As the $\nu_{\pm,dyn}^{\mathrm{3rd}}$ and $\nu_{\pm,dyn}^{\mathrm{4th}}$ contain higher order cross-correlations between $N_{+}$ and $N_{-}$, hence they are expected to amplify the dynamical charge fluctuation signal as compared to $\nu_{\pm,dyn}^{\mathrm{2nd}}$. Moreover, these observables are insensitive to statistical fluctuations due to the subtraction of the statistical contributions in Eq.~\ref{Eq:Eqnudyn2},~\ref{Eq:Eq3rdnudyn}, and ~\ref{Eq:Eq4thdyn}. To understand the behaviour of these observables, a simulation study has been performed using an event generator known as Simulating Many Accelerated Strong-interacting Hadrons (SMASH). The detailed description and simulation results are discussed in the following sections.  

\section{SMASH model description}
\label{Sect:SMASHmodel}

 In this paper, a new hadronic transport approach, SMASH ~\cite{SMASH:2016zqf}, has been applied to study the dynamical charge fluctuations in heavy-ion collisions at center of mass energies available at RHIC. This new microscopic transport approach is relevant to provide a better understanding of the late stage evolution of resonance excitations and decays with vacuum properties in heavy-ion collisions. In this model, the conditions for collision geometry are employed as in the UrQMD (Ultra-relativistic Quantum Molecular Dynamics) approach~\cite{Bass:1998ca}.
 For the initial conditions for heavy-ion collisions, the Woods-Saxon distributions of nuclei are considered 
 in coordinate space. The most well-established hadronic states with their corresponding decays and cross sections are implemented in this model. The detailed discussion and validation of this model with different experimental data can be found in ~\cite{SMASH:2016zqf}. The simulation and analysis details are discussed in Sec.~\ref{Sect:SimAna}.\\

\section{Simulation and Analysis details}
\label{Sect:SimAna}
 In this simulation, the Au+Au collisions with approximately two millions events at center of masss energy, \sNN\ = 200, 62.4 , 27 , 19.6 , 14.5, 11.5, 9.2 and 7.7 GeV are generated within impact parameter range between 0 and 16 fm. A fixed time steps option and with the end time at 30 fm/$\it{c}$ are used to simulated these events. The above collisions energies are the part of RHIC beam energy scan (BES) program including phase-I and II.

\subsection{Analysis details}
\label{Sect:Analysis}

 This study is performed based on the procedure applied in the STAR experiment at RHIC. In heavy-ion experiment, the centrality definition is done based on the charged particles multiplicity distribution at mid rapidity. In this study, the charged particles multiplicity distribution within pesudorapidity ($\eta$) range -1.5 < $\eta$ < 1.5 is used to determine the centrality in Au+Au collisions. These centrality classes are determined for different collision energies.  The nine centrality bins based on the fraction of generated events are selected for this study and those are 0-5\%, 5-10\% 10-20\%, 20-30\%, 30-40\%, 40-50\%, 50-60\%, 60-70\% and 70-80\%.  
 
 The event-by-event positive and negative charged particles are counted within  0.2 GeV/$\it c$ < \pT\ < 5.0 GeV/$\it c$ and $|\eta|$ < 1.5. All the stable hadrons in the PDG have been selected in this simulation.  
 
 To avoid the dependence on the central bin width, the value of the observable is determined using unit bin method. In this method, value of the observable ($\nu$) for each multiplicity bin in centrality class is calculated and then averaged over the width of particular centrality bin with the weights corresponding to relative cross section. The weighted average for the observables are calculated as:
\begin{equation}
 \nu (J_{min} < j < J_{max}) =  \frac{\sum \nu (j)w(j)}{\sum w(j)}.
\end{equation}
 Here, $w(j)$ is the weight of $j^{th}$ multiplicity bin; the $J_{min}$ and $J_{max}$ represent the minimum and maximum range of multiplicity of each centrality class. The $\nu$ observable can be of any order as mentioned in Eq.~\ref{Eq:Eqnudyn2}, ~\ref{Eq:Eq3rdnudyn}, and~\ref{Eq:Eq4thdyn}. 

\subsection{Statistical uncertainty calculation}
\label{Sect:StatError}
 As discussed in Sec.~\ref{Sect:MotivAndObser}, the different orders of dynamical charge fluctuations observable are less sensitive to statistical fluctuation, hence it is expect to have significantly smaller statistical uncertainty due to the presence of some residual statistical fluctuations. To calculate the statistical uncertainty, the Bootstrap method is used and this method is widely used in the event-by-event fluctuation measurements in heavy-ion collisions~\cite{STAR:2014egu,STAR:2013gus, STAR:2020tga}. In this method, seventy Bootstrap samples are used to calculate the standard deviation.


\section{Results and Discussions}
\label{Sect:resultsAndDis}

\indent The higher order dynamical charge fluctuations measures, \nudynM, provide information about the strength of higher order particle correlations which is subjected to change with the total multiplicity of particles in an event ensemble. These observables can be used to study the charge fluctuations developed at the time of phase transition in the heavy-ion collisions. In this section, the behaviour of these observables, in Eq.~\ref{Eq:Eqnudyn2}, ~\ref{Eq:Eq3rdnudyn}, and~\ref{Eq:Eq4thdyn}, with centrality and at different collision energies have been discussed.

\begin{figure}
        \centering
         \scalebox{0.75}
       { \includegraphics{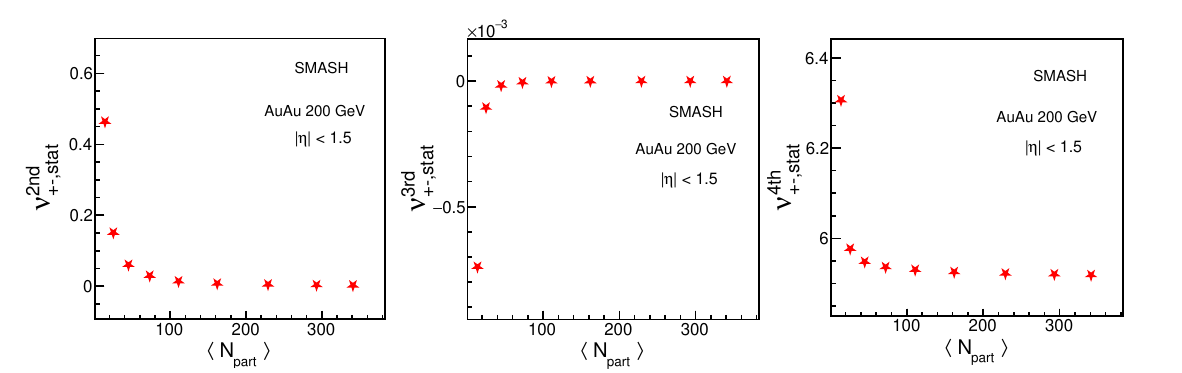}}
        \caption{ The \nuStatTwo (left) , \nuStatThree (middle), \nuStatFour (right) for the charge fluctuations in pseudo-rapidity range $|\eta| < 1.5$ for Au+Au collisions at $\sqrt{s_{NN}}$ = 200 GeV as a function of number of participating nucleons.}
        \label{Fig:StatFluct}
    \end{figure}

Fig.~\ref{Fig:StatFluct} shows the different order of statistical fluctuations contributing to their respective dynamical charge fluctuations at \sNN=200 GeV in Au+Au collisions. The expressions of \nuStatTwo, \nuStatThree, and \nuStatFour\ can be found in Sec.~\ref{Sect:MotivAndObser}. It shows \nuStatTwo\ and \nuStatFour\ are positive and behave the same way unlike \nuStatThree. The \nuStatThree\ is negative as the total number of positively charged particles larger than the negatively charged particles in heavy-ion collisions. The values are significantly changed at peripheral collisions as the $N_{+}$ and $N_{-}$ change rapidly.  These statistical contributions are subtracted in order to get the dynamical charge fluctuations as discussed below.  

    \begin{figure}
        \centering
        \scalebox{0.75}
        {\includegraphics{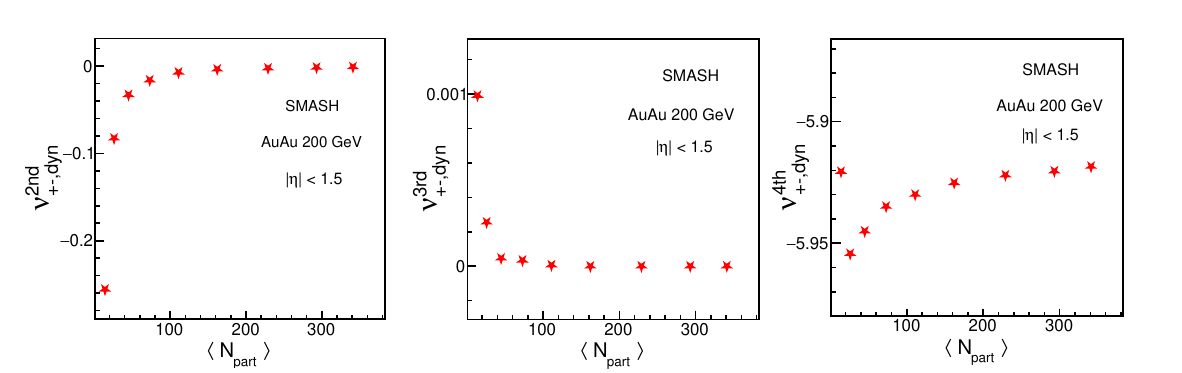}}
        \caption{The dynamical charge fluctuations, \nudynTwo (left) , \nudynThree (middle), \nudynFour (right) for the charged particles measured in pseudo-rapidity range $|\eta| < 1.5$ for Au+Au collisions at $\sqrt{s_{NN}}$ = 200 GeV as a function of number of participating nucleons.}
         \label{Fig:dynauau200}
    \end{figure}

Fig.~\ref{Fig:dynauau200} shows the value of \nudynTwo, \nudynThree, and \nudynFour for different centrality at \sNN = 200 GeV in Au+Au collisions.  The statistical uncertainties are smaller than the marker size.  The values of \nudynTwo\ and \nudynFour\ are negative whereas \nudynThree\ is positive finite at all centrality. In the higher orders \nudynM\ , the higher order correlation terms dominate as discussed in Sec.~\ref{Sect:MotivAndObser}. 

\begin{figure}
        \centering
        \scalebox{0.75}
        {\includegraphics{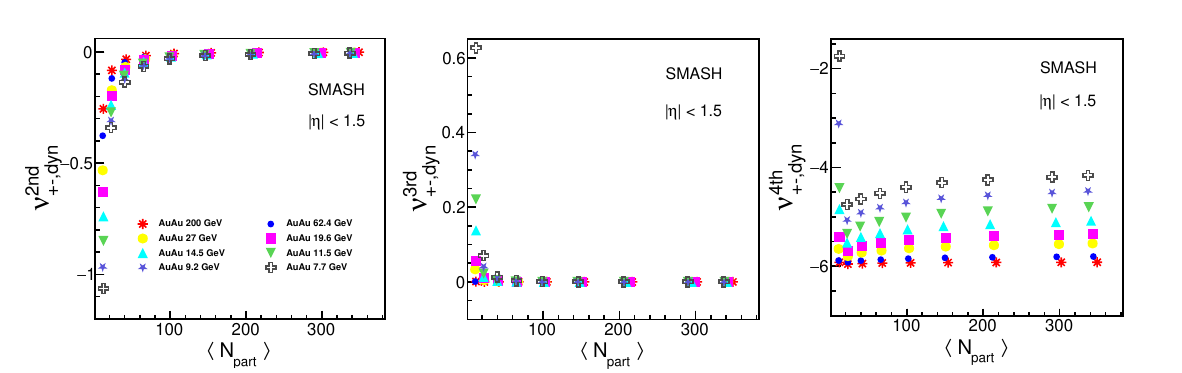}}
        \caption{The dynamical charge fluctuations, \nudynTwo (left) , \nudynThree (middle), \nudynFour (right) for the charged particles measured in pseudo-rapidity range $|\eta| < 1.5$ for Au+Au collisions at $\sqrt{s_{NN}}$ = 200 GeV to 7.7 GeV as a function of number of participating nucleons. }
        \label{Fig:CentDyn}
    \end{figure}

Fig.~\ref{Fig:CentDyn} shows these three observables for the eight collisions energies between \sNN = 200 to 7.7 GeV in Au+Au collisions. At a given centrality, they show a clear energy dependence. The values of \nudynThree\ are very small and show no significant difference between different collisions energies. This could be due to \nudynThree\ is less sensitive to charge fluctuations in heavy-ion collisions in the SMASH model. This would be important to investigate this feature in the data. On the other hand, the \nudynFour\ shows a noticeable difference at different collision energies relative to its lower order. The fourth order \nudynFour\ is relatively more sensitive to charge fluctuations due to presence of higher order correlation terms. Therefore, in heavy-ion experiments it can provide vital information about the phase transition. 

\begin{figure}
        \centering
        \scalebox{0.75}
        {\includegraphics{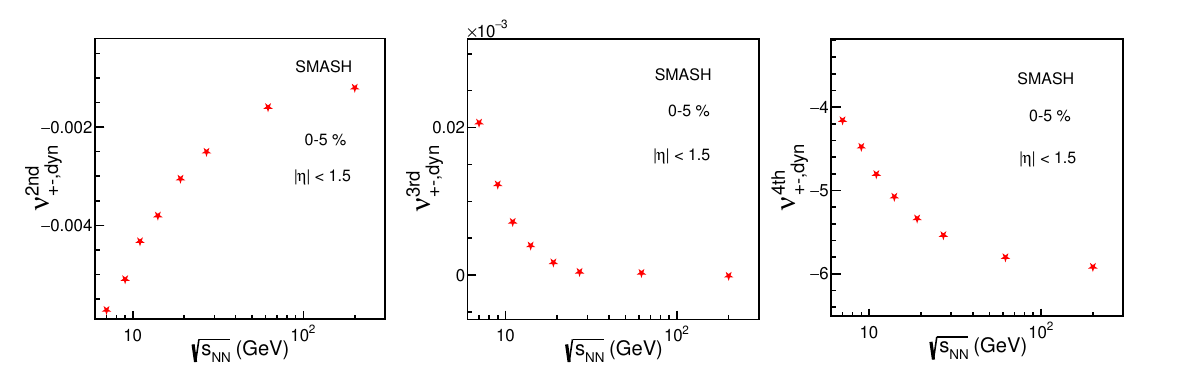}}
        \caption{ The dynamical charge fluctuations, \nudynTwo (left) , \nudynThree (middle), \nudynFour (right) as a function beam energy for 0-5\% centrality in Au+Au
        collisions.}

        \label{Fig:EngDepDynChr}
    \end{figure}

Fig.~\ref{Fig:EngDepDynChr} shows the collision energy dependence of the \nudynTwo, \nudynThree, \nudynFour\ in 0-5\% central Au+Au collisions using the SMASH model. The \nudynTwo\ shows a increasing trend with increasing collision energy whereas \nudynThree\ and \nudynFour\ show a decreasing trend with increasing collision energy. 
   
These higher order dynamical charge fluctuation measures can shed light on the parton-to-hadron phase transition in heavy-ion experiments. These observables are measured in finite acceptance and centrality, so the volume fluctuations could play a significant role. The effect of global charge conservation in a finite acceptance also show an importance as discussed in ~\cite{Sharma:2015lva}. Furthermore, the resonance decay contribution could contribute the higher order \nudynM. It needs more study of these effects and possible impact on these observables in heavy-ion experiments and beyond the scope of this paper. Although similar discussion can be found for \nudynTwo\ in ~\cite{Jeon:2000wg,Bleicher:2000ek}.  This is an attempt to explore new observable that is sensitive to charge fluctuations and amplify the signal in heavy-ion collisions experiments at RHIC and the LHC. A detailed theoretical work on these observables is needed to understand the hardon gas and QGP limit. 
     
\section{Summary and outlook}
\label{Sect:SummaryOutlook}
This is an attempt to introduce the higher order charge fluctuation observable those are sensitive to the quark-to-hadron phase transition and can amplify the signal in heavy-ion collisions. This work is augmentation of previous work on the dynamical charge fluctuation measure that is second order, \nudynTwo\cite{Sharma:2015lva}. In this paper we introduce the third and fourth order dynamical charge fluctuations and provide the expression of \nudynThree\ and \nudynFour. These observables contain the higher order cross-correlation terms, hence they can amplify the signal in heavy-ion collisions. 

The SMASH model is used to study the behavior of the \nudynTwo, \nudynThree, and \nudynFour\ as function of centrality and collision energies. It is observed that the \nudynFour\ can amplify the signal compared to \nudynTwo\ and \nudynThree. It is also found that the \nudynThree\ and \nudynFour\ are sensitive to detector effects unlike the \nudynTwo. 

Various effects like the volume fluctuations, the global charge conservation, and the resonance effects may play a role which need to be studied as an outlook of this work. Furthermore, theoretical calculation pertaining to these observables are required to set the limit of hadron gas and partonic phase of matter. The measurements at the LHC and RHIC would be interesting to see the signature of phase transtion in heavy-ion experiments using these observables.

\subsection*{\textbf{Acknowledgement}}
BS is thankful to  Nihar Ranjan Sahoo for helpful discussions and suggestions that initiated this work. She is also very grateful to Tapan Kumar Nayak, Bedangadas Mohanty and Subhasish Chattopadhyay for their support. This research used resources of the LHC grid computing center at the Variable Energy Cyclotron Center, India. The simulations and computations were also supported in part by the SAMKHYA: High Performance Computing Facility provided by the Institute of Physics, Bhubaneswar.

\end{document}